\documentclass[12pt]{article}
\newcommand{\x}{arXiv:}

\usepackage[nosort]{cite}
\usepackage{graphicx}
\usepackage[font={scriptsize,singlespacing}]{caption}
\usepackage{epstopdf}
\usepackage{amsmath}
\usepackage{amssymb}
\usepackage{subfigure}
\usepackage{hyperref}
\usepackage{url}
\usepackage{xcolor}
\usepackage{color}

\def\sideremark#1{\ifvmode\leavevmode\fi\vadjust{\vbox to0pt{\vss% the remark
 \hbox to 0pt{\hskip\hsize\hskip1em%                          will appear only
 \vbox{\hsize2cm\tiny\raggedright\pretolerance10000%          on the side
 \noindent #1\hfill}\hss}\vbox to8pt{\vfil}\vss}}}%
\definecolor{amaranth}{rgb}{0.9, 0.17, 0.31}
\definecolor{purple(munsell)}{rgb}{0.62, 0.0, 0.77}
\definecolor{americanrose}{rgb}{1.0, 0.01, 0.24}
\definecolor{palatinateblue}{rgb}{0.15, 0.23, 0.89}
\definecolor{royalblue(web)}{rgb}{0.25, 0.41, 0.88}
\definecolor{hanpurple}{rgb}{0.32, 0.09, 0.98}
\definecolor{beaublue}{rgb}{0.74, 0.83, 0.9}
\definecolor{carminered}{rgb}{1.0, 0.0, 0.22}
\definecolor{brightpink}{rgb}{1.0, 0.0, 0.5}
\hypersetup{ linktoc=all,
    colorlinks, linkcolor={palatinateblue},
    citecolor={brightpink}, urlcolor={purple(munsell)}
}

\begin{document}
\thispagestyle{empty}
\begin{center}

\null \vskip-1truecm \vskip2truecm

{\Large{\bf \textsf{Never Judge a Black Hole by Its Area}}}

\vskip1truecm

\textbf{\textsf{Yen Chin Ong}}\\
\vskip0.3truecm
{Nordita, KTH Royal Institute of Technology and Stockholm University, \\ Roslagstullsbacken 23,
SE-106 91 Stockholm, Sweden}\\
{\tt Email: yenchin.ong@nordita.org}\\

\end{center}
\vskip1truecm \centerline{\textsf{ABSTRACT}} \baselineskip=15pt

\medskip

Christodoulou and Rovelli have shown that black holes have large interiors that grow asymptotically linearly in advanced time, and speculated that this may be relevant to the information loss paradox. We show that there is no simple relation between the interior volume of an arbitrary black hole and its horizon area. That is, the volume enclosed is not necessarily a monotonically increasing function of the surface area. 

\addtocounter{section}{1}
\section* {\large{\textsf{1. Black Holes and Their Large Interiors}}}

An asymptotically flat Schwarzschild black hole has a spherical event horizon. The usual metric of this geometry in 4-dimensions, in the units $G=c=1$, is
\begin{equation}
g[\text{Sch}]=-\left(1-\frac{2M}{r}\right)dt^2 + \left(1-\frac{2M}{r}\right)^{-1} dr^2 + r^2\left(d\theta^2 + \sin^2\theta d\phi^2 \right),
\end{equation}
where $M$ is the ADM mass, and $r$ is the areal radius. That is, the area of the event horizon $r_h$ is the area of the 2-sphere: $4\pi r_h^2$. 

Unlike its surface area, the concept of ``the volume'' of a black hole is a tricky one. The reason is that volume depends on the choice of 3-dimensional spacelike hypersurface. As such, no unique volume can be prescribed to a black hole.  Furthermore, the interior of a static black hole is nevertheless dynamical, so one should definitely \emph{not} think of a black hole as a black box that bounds a certain amount of volume that can be easily estimated from knowing the size of its area. This is well-known: a maximally extended Schwarzschild [Kruskal-Szekeres] geometry has an infinitely large asymptotically flat region on the ``other side'', connected via the Einstein-Rosen bridge.  Similarly, one could attach a closed FLRW universe to the interior of a black hole via the Einstein-Rosen bridge, resulting in the so-called Wheeler's ``bag-of-gold'' geometry \cite{Wheeler}. 
Even non-black hole configurations can have arbitrarily large interiors than their areas might suggest \cite{monsters}.
What about black holes that were formed from gravitational collapse and have no second asymptotic region? 

One important motivation to study the interiors of \emph{generic} black holes is of course the information loss paradox. As matter falls into a black hole and the black hole gradually Hawking evaporates away, it seems that when the black hole disappears, the information of the in-fallen matter will be lost forever, thereby threatening the unitarity of quantum mechanics. One proposal to resolve this paradox is the idea that black holes do not completely evaporate away, but instead settle down to a [meta-]stable remnant. [See \cite{1412.8366} for a recent review of black hole remnants.] An obvious shortcoming of such a proposal is that, as the black hole shrinks in size it would seem that room is running out for storing a large amount of information. This objection presumably does not arise if the interior of a rather small black hole can remain large\footnote{The usual objection against black hole remnants is that they will be copiously produced [``infinite production problem'']. Large interiors may ameliorate this \cite{0901.3156}.}.  However, since ``bag-of-gold'' geometry is unlikely to be generic, and Kruskal-Szekeres geometry only holds for eternal black holes, we have to look elsewhere to resolve the information loss paradox, which is most important in the case of black holes formed from gravitational collapse. Fortunately, it turns out that even such black holes have large interiors. [Whether or not such interior volumes remain sufficiently large even as the black hole shrinks to Planck scale is, of course, a relevant and important question.]

In a recent paper, Christodoulou and Rovelli \cite{1411.2854} pointed out that, while there is no unique volume that can be prescribed to the interior of a black hole, one could look at the volume of the \emph{largest} spacelike spherically-symmetric surface bounded by the event horizon of the black hole. Such a volume is a geometrical property that is coordinate-independent. Christodoulou and Rovelli [hereinafter, CR] showed that most of the volume contribution comes from a region which is not causally connected with matter that has fallen far into the black hole [see the work of Bengtsson and Jakobsson \cite{1502.01907} for an explicit and nice illustration of this fact, as well as their generalization of the work of CR to the case of asymptotically flat Kerr black holes.] We will henceforth refer to such a volume measure as the ``CR-volume''.

For an asymptotically flat Schwarzschild black hole in 4-dimensions, most of the CR-volume contribution is given by the integral \cite{1411.2854, 1502.01907}
\begin{equation}\label{int1}
\text{Vol.} \sim \int^v \int_{S^2}  \max\left[r^2 \sqrt{\frac{2M}{r}-1}\right]  \sin\theta ~d\theta d\phi dv,
\end{equation}
where $v$ is the advanced time defined by
\begin{equation}
 v := t + \int \frac{dr}{f(r)} = t + r + 2M \ln |\frac{r}{2M}-1|; ~~f(r):=1-\frac{2M}{r}.
\end{equation}
Following \cite{1502.01907}, we omitted the lower limit in the integral with respect to $v$, which only contributes to a negligible finite value, whereas the integral will be dominated by its upper limit $v$. The coefficient of $v$ can be maximized by maximizing the function
\begin{equation}
 \mathcal{F}(r) := r^2 \sqrt{\frac{2M}{r}-1}.
\end{equation}
Elementary calculus shows that $r=3M/2$ maximizes $\mathcal{F}(r)$. 
Indeed, most of the volume comes from the contribution of this constant $r$ slice. [Note that this slice is rather ``close'' to the event horizon $r=2M$.]
This leads to 
\begin{equation}
\text{Vol.} \sim  3\sqrt{3} \pi M^2 v.
\end{equation}
That is, the CR-volume grows asymptotically linearly in $v$. In other words, even though a static black hole looks the same to the exterior observer no matter how long one waits [this is a classical statement without taking into account Hawking radiation], its interior gets larger with time [c.f. another proposal of black hole volume in \cite{0508108}]. 

The estimate given by CR is that the supermassive black hole at the center of our galaxy, Sagittarius A$^*$, contains sufficient space to fit a \emph{million} solar systems, despite its areal radius is only a factor of 10 or so larger than the Earth-Moon distance. 
Taking into account the rotation of the black hole does not change this result by much, despite the rotation rate of Sagittarius A$^*$ is about 90\% of the extremal limit. In other words, the CR-volume for asymptotically flat black holes seem to be robust against rotational effect, as long as it stays below $\sim 99\%$ of the extremal limit \cite{1502.01907}.  

In view of the potentially important role the CR-volume may play in the resolution of the information loss paradox [as suggested by CR \cite{1411.2854}], the properties of the CR-volume of other black hole solutions should be investigated. For example, one may wish to consider black holes with other asymptotic geometries.  One especially interesting arena to explore is the anti-de Sitter [AdS] space --- in the presence of a negative cosmological constant, a large class of black hole solutions are allowed. Unlike their asymptotically flat cousins, these black holes can have non-trivial horizon topologies, hence are often referred to as ``topological black holes'' \cite{9705004, 9705012, 9709039, 9808032}. The metric tensor of a static topological black hole in $(n+2)$-dimensions [$n\geqslant 2$] is given by:
\begin{flalign}\label{metric}
g[\text{AdSRN}^k_{n+2}] =& -\left[k+\frac{r^2}{L^2}-\frac{16\pi M}{n \Gamma[X^k_n]r^{n-1}} + \frac{8\pi Q^2}{n(n-1)(\Gamma[X^k_n])^2r^{2n-2}}\right]dt^2 \\ & \notag + \left[k+\frac{r^2}{L^2}-\frac{16\pi M}{n \Gamma[X^k_n]r^{n-1}} + \frac{8\pi Q^2}{n(n-1)(\Gamma[X^k_n])^2r^{2n-2}}\right]^{-1} dr^2 + r^2 d\Omega^2 [X^k_n],
\end{flalign}
where $L$ is the AdS curvature length scale, and $d\Omega^2[X^k_n]$ is a metric of constant curvature $k=\left\{-1,0,+1\right\}$ on the $n$-dimensional Riemannian manifold $X^k_n$, with [dimensionless] area $\Gamma[X^k_n]$. For example, if $X^k_n$ is a 2-sphere, then $\Gamma[X^k_n]=\Gamma[S^1_2] = 4\pi$. 
Note that in general, the geometry is only asymptotically \emph{locally} AdS.

\addtocounter{section}{1}
\section* {\large{\textsf{2. The Interiors of Flat AdS Black Holes}}}

In the $k=0$ case, the metric Eq.(\ref{metric}) reduces to the simpler form:
\begin{flalign}\label{flatmetric}
g[\text{AdSRN}^0_{n+2}] =& -\, \Bigg[{r^2\over L^2}\;-\;{16 \pi M^*\over n r^{n-1}}+{8\pi Q^{*2}\over n(n-1) r^{2n-2}}\Bigg]dt^2\;  \\ \notag &+\;\left[{r^2\over L^2}\;-\;{16 \pi M^*\over nr^{n-1}} +{8\pi Q^{*2}\over n(n-1) r^{2n-2}}\right]^{-1}dr^2 \;+\;r^2\Bigg[\sum_{i=1}^n d\zeta_i^2\Bigg],
\end{flalign}
where the $\zeta_i$'s are dimensionless coordinates on a flat space.  
If the black hole has a compact horizon with toral topology [$T^n = \Bbb{R}^n/\Bbb{Z}^n$],  the mass parameter $M^*$ is simply $M/(2\pi K)^n$, where $K$ is a dimensionless ``compactification parameter''. Similarly  $Q^* = Q/(2\pi K)^n$. For example, if the horizon is a flat 2-torus, one may think of it as a product of two circles: $T^2 = S^1 \times S^1$, each of which has period $2\pi K$, so the dimensionless area corresponds to the event horizon is $\Gamma[X^k_n=T^0_2]=4\pi^2 K^2$. The dimensionful area is $4\pi^2K^2 r_h^2$, where $r_h$ denotes the horizon. This generalizes straightforwardly to $(2\pi K)^n r_h^n$ for general dimensions $n \geqslant 2$.

Of course, even if we restrict to the case in which the horizon is compact, there are still many other possible topologies. In 5-dimensions, i.e., if the event horizon is a 3-dimensional manifold, there are 6 orientable compact topologies with constant zero curvature \cite{0311476}, namely the torus $T^3$ and its various quotient topologies: the dicosm $T^3/\Bbb{Z}_2$, the tricosm $T^3/\Bbb{Z}_3$, the tetracosm $T^3/\Bbb{Z}_4$, the hexacosm $T^3/\Bbb{Z}_6$, and the didicosm [also called the Hantzsche-Wendt space] $T^3/(\Bbb{Z}_2 \times \Bbb{Z}_2)$. In such cases, we still \emph{define} the area as $8\pi^3K^3$. That is, we use $K$ to measure the relative size of these spaces.   
We can also consider a non-compact [planar] event horizon. To do this, we take $M$, $Q$ and $K$ to infinity, but in a way such that the parameters $M^*$ and $Q^*$ remain finite.

Let us consider for concreteness,  toral black holes in 4-dimensional spacetime. These flat black holes exhibit many surprisingly elegant properties \cite{1403.4886}, such as:
\begin{itemize}
\item[(1)] The maximal in-falling time $\tau_{\text{max}}$ from the horizon to the singularity for a neutral toral black hole\footnote{For an asymptotically flat Schwarzschild black hole, we have instead \begin{equation} \tau_{\text{max}}= \int_0^{2M} \left(\frac{2M}{r} -1 \right)^{-\frac{1}{2}}dr = \pi M. \notag \end{equation}} is independent of the black hole mass $M$:
\begin{equation}\label{wow}
\tau_{\text{max}} = \int_0^{r_h} \left(\frac{2M}{\pi K^2r} -\frac{r^2}{L^2}\right)^{-\frac{1}{2}} dr = \frac{\pi L}{3}, ~~r_h = \left(\frac{2ML^2}{\pi K^2}\right)^{\frac{1}{3}}.
\end{equation}
\item[(2)] In the charged case, the Kretschmann scalar $R_{abcd}R^{abcd}$ at the event horizon of the extremal black hole is precisely $144/L^4$. [It is also the square of the scalar curvature $R=-12/L^2$.] In the neutral case, the Kretschmann scalar at the event horizon is $36/L^4$. In both of these cases,  the Kretschmann scalar at the horizon is independent of the black hole mass, and only depends on $L$.
\end{itemize}
In some sense, these black holes are thus behaving more like pure AdS space than black holes [property (1) should be compared to the fact that in pure AdS, the time to fall from anywhere to the ``center'' of AdS only depends on the curvature radius]. We thus expect some surprises in the properties of the CR-volume of these black holes. As we shall see below, this is indeed so.

As pointed out by CR \cite{1411.2854}, their analysis generalizes in a straight-forward manner to other spherically symmetric black holes. 
In the case of neutral toral black holes in 4-dimensions,
the metric is
\begin{equation}
g[\text{AdSRN}^0_{4}] = -\left(\frac{r^2}{L^2}-\frac{2M}{\pi K^2 r}\right)dt^2 + \left(\frac{r^2}{L^2}-\frac{2M}{\pi K^2 r}\right)^{-1}dr^2 + r^2(d\zeta^2 + d\xi^2); ~~\zeta, \xi \in [0, 2\pi K).
\end{equation}

Its CR-volume is
\begin{equation}\label{v1}
\text{Vol.} \sim \int^v  \int_{T^2}  \max\left[r^2 \sqrt{\frac{2M}{\pi K^2 r} -\frac{r^2}{L^2}}\right]~  d\zeta d\xi dv.
\end{equation}
The coefficients in front of $v$ is maximized when 
\begin{equation}
r=r_*=\left(\frac{ML^2}{\pi K^2}\right)^{\frac{1}{3}}.
\end{equation}
Note that $r_* < r_h$. 
Substituting the value of $r_*$ into the function
\begin{equation}
\mathcal{F}(r) =  r^2 \sqrt{\frac{2M}{\pi K^2 r} -\frac{r^2}{L^2}}, 
\end{equation}
we get $\mathcal{F}(r_*)=ML/(\pi K^2)$.

We therefore find that the CR-volume is
\begin{equation}
\text{Vol.} \sim 4\pi^2 K^2 \left[\frac{ML}{\pi K^2}\right] v = 4\pi ML v,
\end{equation}
which is independent of the compactification parameter $K$.
Alternatively, we can show this by checking that 
\begin{equation}
\frac{\partial}{\partial K} \left[4\pi K^2 \mathcal{F}(r_*) \right]= 0.
\end{equation}

The result generalizes to arbitrary dimension. The value of $r$ that maximizes the volume is
\begin{equation}
r_* = \left[\frac{8\pi ML^2}{n (2\pi K)^n}\right]^{\frac{1}{n+1}}, 
\end{equation}
The CR-volume can be shown to be 
\begin{equation}\label{CR1}
\text{Vol.} \sim (2\pi K)^n \mathcal{F}(r_*) v = \frac{8\pi ML}{n}v.
\end{equation}

Thus, in arbitrary dimension, the CR-volume of a neutral AdS black hole with toral event horizon is \emph{independent} of the compactification parameter. This is surprising since our intuition of volume would be that larger surface area can bound a larger interior volume. This is certainly still true for the CR-volume of an asymptotically flat Schwarzschild black hole --- for 4-dimensional case, we have  $\text{Vol.}\sim 3\sqrt{3} \pi M^2 v = 3\sqrt{3}Av/16$, where $A=4\pi r_h^2 = 16 \pi M^2$ is the area of its horizon. Thus a larger Schwarzschild black hole \emph{does} have a larger interior volume\footnote{Here, as well as later, we are comparing the CR-volumes at the same [sufficiently large] constant advanced time $\nu$.}, which agrees with our flat space expectation.  This is perhaps only a coincidence. As we have seen, the flat AdS black holes behave in an entirely different way --- for fixed mass $M$, the CR-volume of a black hole with a larger horizon\footnote{In 4-dimensions, the exact expression for the horizon area is $\displaystyle 4\pi^2K^2\left(\frac{2ML^2}{\pi K^2}\right)^{2/3} \propto K^{2/3}$.} [i.e. with larger $K$] grows asymptotically the same way as a black hole with a smaller horizon [i.e. with smaller $K$]. In other words, at sufficiently late time, the maximum volume in any flat AdS black hole with compact event horizon is identical [up to some finite pieces which are negligible in the volume integral].  
This provides an example in which a large black hole can nevertheless have small interior volume.
Note that the planar case is different since there we have to take both $M$ and $K$ to infinity, and the expression given in  Eq.(\ref{CR1}) does in fact diverge.

This remarkable behavior of $K$-independence of the CR-volume nevertheless only holds if the black holes are neutral\footnote{CR discussed the interior volume of asymptotically flat Reissner-Nordstr\"om black holes\cite{1411.2854}.}. The proof is straightforward but tedious. 
For simplicity, let us consider only the 4-dimensional case.
One starts with the function to be maximized:
\begin{equation}
\mathcal{F}(r) = r^2 \sqrt{\frac{2M}{\pi K^2 r}-\frac{r^2}{L^2} - \frac{Q^2}{\pi K^2 r^2}}.
\end{equation}
Let us denote the value of $r$ that maximizes $\mathcal{F}(r)$ by $r_*$. This can be solved analytically but it is extremely complicated. Fortunately we do not need the exact expression of $r_*$. It can be shown by elementary calculus that $r_*$ satisfies the equation
\begin{equation} \label{condition}
3\pi K^2 r_*^4 - 3ML^2 r_* + L^2Q^2 = 0. 
\end{equation}
Using this condition, we can re-write
\begin{equation}
\mathcal{F}(r_*)  = \frac{r_*}{\sqrt{\pi}KL}\left[\pi K^2 r_*^4 - \frac{Q^2L^2}{3}\right]^{\frac{1}{2}}.
\end{equation}
The CR-volume is
\begin{equation}
\text{Vol.} \sim 4\pi^2 K^2 \mathcal{F}(r_*) v.
\end{equation}
Therefore to check for $K$-dependence, it suffices to check whether
\begin{equation}
\frac{\partial}{\partial K} \left[K^2 \mathcal{F}(r_*)\right] = 0.
\end{equation}
A straightforward algebraic manipulation, using the fact that from Eq.(\ref{condition}),
\begin{equation}
\frac{\partial r_*}{\partial K} = \frac{6\pi K r_*^5}{L^2Q^2 - 9\pi K^2 r_*^4},
\end{equation}
yields the result that 
\begin{equation}
\frac{\partial}{\partial K} \left[K^2 \mathcal{F}(r_*)\right] = 0 \Longleftrightarrow Q=0.
\end{equation}
Therefore, only neutral toral black holes can have $K$-independent CR-volumes. 

\addtocounter{section}{1}
\section* {\large{\textsf{2. The Interiors of AdS Black Lenses}}}

We have discussed the intriguing property of the CR-volume for neutral toral black holes in AdS, namely that it is independent of the compactification parameter $K$.
It is therefore of paramount importance to compare this result with AdS black holes with positively curved event horizons. There is an anologous ``compactification parameter'' one can consider here, namely by taking quotients of the spherical horizons. More specifically, one constructs the quotient space $S^3/G$ of the 3-sphere by a finite subgroup $G$ of  $\text{SO}(4)$ acting freely on $S^3$. 
By the classification theorem of closed surfaces, there is no other orientable topology other than $S^2$ for a positively curved 2-dimensional compact horizon. So let us consider horizons with $S^3$ topology in 5-dimensional spacetime instead. We can then construct the so-called ``black lens'' \cite{ida}, which is also allowed in asymptotically flat spacetime [a rotating black lens was constructed in \cite{0808.0587}].

A black lens is a black hole with lens space topology $L(p,1):=S^3/\Bbb{Z}_p$, where $p$ is a positive integer\footnote{More generally, a lens space $L(p,q)$ is characterized by two parameters. It is the 3-manifold obtained by gluing the boundaries of two solid tori together such that the meridian of the first goes to a $(p,q)$-curve on the second. By $(p,q)$-curve we mean a smooth curve that wraps around the meridian $q$ times and around the longitude $p$ times. A lens space can also be constructed by making appropriate identifications on the boundary 2-sphere of a solid ball. See, e.g., \cite{maunder}.}. 
For example, 
$S^3/\Bbb{Z}_1$ is just $S^3$ itself. The first non-trivial example is $S^3/\Bbb{Z}_2 \cong \Bbb{R}\text{P}^3$, the real projective 3-space. Note that $\Bbb{R}\text{P}^n$ is orientable when $n$ is odd. The parameter $p$ is, in some sense, analogous to the compactification parameter $K$ for the flat AdS black holes\footnote{Of course, the lens space is not the only possible topology; one could take quotients by other groups. If we assume the geometry is homogeneous, then in addition to the cyclic groups $\Bbb{Z}_p$ of order $p$, one can take $G$ to be the binary dihedral groups of order $4n$, where $n \geqslant 2$; the binary tetrahedral group of order 24; the binary octahedral group of order 48; and the binary icosahedral group of order 120 \cite{wolf}.}. Of course, $p$ is discrete while $K$ is continuous. Nevertheless, it would be interesting to see what happens to the CR-volume of black lenses as we take $p$ to infinity.

The metric of a black lens is given by Eq.(\ref{metric}), with $k=1$ and $X^1_3 = S^3/\Bbb{Z}_p$. 
The area of $S^3$ is $2\pi^2$, so the area of  $S^3/\Bbb{Z}_p$ is simply $2\pi^2/p$. The area of the horizon is therefore given by
\begin{equation}
\frac{2\pi^2}{p} r_h^3 = \frac{12^{1/4}\pi^{5/4}L^{3/2} (\sqrt{3\pi L^2 + 32pM}-L\sqrt{3\pi})^{3/2} }{6p}.
\end{equation}
For fixed $M$ such that $M$ is small\footnote{These are ``small'' black holes that are subjected to Hawking evaporation, unlike the ``large'' black holes that attain thermal equilibrium with their own Hawking radiation.} compared to $L^2$, 
this function initially \emph{increases} with $p$ but eventually reaches a maximum, and then starts to decrease with increasing $p$. That is to say, for $M$ small compared to $L^2$,
taking quotient with a sufficiently small group $\Bbb{Z}_p$ actually causes the area of the event horizon to \emph{increase}.
However, for fixed $M$ such that $M$ is sufficiently large compared to $L^2$, the horizon area always decreases upon taking quotients. [See also the discussion in \cite{0806.3818} for an expression of the horizon area in terms of the dimensionless area $\Gamma[L(p,1)]$.]
Thus in either case, for sufficiently large $p$, the black hole horizon is small. 

To compute the CR-volume, the function that we need to maximize is now 
\begin{equation}
\mathcal{F}(r) = r^3 \sqrt{\frac{16\pi M}{3r^2\left(\frac{2\pi^2}{n}\right)}-\frac{r^2}{L^2}-1}=r^3\sqrt{\frac{8nM}{3\pi r^2}-\frac{r^2}{L^2}-1}.
\end{equation}
It turns out that $\mathcal{F}(r)$ is maximized by
\begin{equation}
r=r_* = \sqrt{\frac{3L^2}{8}\left(-1+\sqrt{1+\frac{256 p M}{27 \pi L^2}}\right)}, 
\end{equation}
and the CR-volume is
\begin{flalign}\label{v2}
\text{Vol.} &\sim \frac{2\pi^2}{p} \mathcal{F}(r_*) v  \\ \notag
& = \frac{2\sqrt{2}\pi Lv}{192 p} \left(\sqrt{27 \pi L^2 + 256 Mp}-3\sqrt{3}L\right) \\ \notag & ~~~~ \times \sqrt{128Mp + 9\pi L^2 - \sqrt{3\pi}\sqrt{27\pi L^2 + 256 Mp}L}.
\end{flalign}
It can be checked that for all values of fixed $M$ and $L$, this expression is monotonically \emph{increasing} in $p$, and in the limit $p \to \infty$, the expression in front of $v$ in the CR-volume tends toward a constant $8\pi MLv/3$. Note that this is the same value as the CR-volume of a 5-dimensional neutral toral black hole [independent of $K$]. It would be interesting to investigate if there is a deeper reason to this remarkable fact.

As we have mentioned, for $M$ that is smaller than $ L^2$, the effect of taking quotient by a small cyclic group $\Bbb{Z}_p$ is to increase the horizon area. Since the CR-volume is monotonically increasing,  these black holes conform to our flat space intuition that the volume is a monotonically increasing function of its area. However if $p$ is sufficiently large, then regardless of the value of the ratio $M/L^2$, the horizon area is monotonically decreasing in $p$, whereas the CR-volume is monotonically increasing in $p$. Here we have an example in which a very small black hole could nevertheless possess much larger interior than a bigger black hole of the same mass.

\addtocounter{section}{1}
\section* {\large{\textsf{3. Conclusion: No Simple Relation between Area and Volume}}}

Despite the fact that no unique volume can be prescribed to a black hole, Christodoulou and Rovelli \cite{1411.2854} showed that it makes sense to talk about the volume of the largest hypersurface bounded by the event horizon [``CR-volume'']. Specifically they investigated the CR-volumes of asymptotically flat Schwarzschild and Reissner-Nordstr\"om black holes. Bengtsson and Jakobsson \cite{1502.01907} recently extended the result to asymptotically flat Kerr black holes. Since interior volumes of black holes may play some roles in the context of information loss paradox, it is crucial to understand the properties of CR-volumes for various other black holes. 

Our flat space intuition is that the volume of a closed surface should be a monotonically increasing function of its area. 
In other words, a smaller surface area means the volume enclosed is also small. 
For example, the volume of a 2-sphere is proportional to $A^{3/2}$, where $A$ is its surface area.
The interior volume proposed by Christodoulou and Rovelli does have such a property in the case of asymptotically flat Schwarzschild black hole [there the volume is proportional to $A$ $(\times v)$]. 
That the same remains true in the case of asymptotically flat Kerr black holes can be seen in \cite{1502.01907} --- for a fixed mass $M$, increasing the angular momentum $J=aM$ would decrease the horizon area $8\pi M (M + \sqrt{M^2-a^2})$ and likewise its CR-volume also decreases. [However no such volume arises in the extremal case despite of the nonzero horizon area. The reason is that the region in which the calculation is valid is not present in the extremal case.] The case for asymptotically flat Reissner-Nordstr\"om black hole is very similar.

In this work we study topological black holes with toral and lens space event horizons and found that the CR-volume of a black hole is \emph{not} always a monotonically increasing function of its horizon area. In the toral case, if the black hole is neutral, then the CR-volume is independent of the compactification parameter. In other words, for fixed black hole mass $M$ and fixed AdS length scale $L$, the CR-volume is independent of the size of the horizon area. 
The CR-volume for an AdS black hole in 5-dimensional spacetime with lens space topology $S^3/\Bbb{Z}_p$ [``black lens''] is even more remarkable --- for any fixed $M$ and $L$, it is monotonically increasing with $p$ and tends to a constant $8\pi MLv/3$ in the limit $p \to \infty$, whereas in the same limit the horizon area shrinks toward 0. 
Thus a smaller black hole can have a larger CR-volume than a bigger hole of the same mass. [We remind the readers that this statement is not exact since the CR-volumes we calculated are asymptotic expressions --- they ignore the lower limit of the integral in, e.g, Eq.(\ref{int1}). ]

Therefore, there appears to be no simple relation between the area of the event horizon and the CR-volume. In particular, the CR-volume certainly is \emph{not} always a monotonically increasing function of its horizon area. It may be interesting to investigate the behavior of the CR-volumes for other black holes, such as black ring, and black hole solutions of modified gravity theories. 

We now speculate on the implication of this result to the information loss paradox. As discussed by Christodoulou and Rovelli, the large interior of a black hole may store information even if its area is shrinking. Of course, this would mean that the Bekenstein-Hawking entropy does not measure the entire information content [i.e., we have to subscribe to the ``weak-form interpretation'' of the Bekenstein-Hawking entropy \cite{1412.8366, 0901.3156, 0003056, 0501103}]. For a generic topological black hole in AdS, its horizon area [and hence its Bekenstein-Hawking entropy] is typically not completely fixed by its mass. For example, for fixed mass $M$ which is small compared to the square of the AdS length scale $L^2$, two AdS black lenses with different values of $p$ can have the same horizon area. If the Bekenstein-Hawking entropy measures the entire information content of a black hole, this implies that these black holes have the same information storage ``capacity''. However, if the weak form interpretation is correct, and the CR-volume is responsible for information storage, then these two black holes would have different capacities. On the other hand, this would mean that for fixed $L$, AdS neutral toral black holes with different horizon areas [due to different values of the compactification parameter $K$] all have the same capacity for information storage, as long as they have the same mass. 

The fact that a neutral AdS \emph{planar} black hole possesses infinite CR-volume is intriguing. Although the Bekenstein-Hawking entropy of a planar black hole is formally infinite, from the perspective of AdS/CFT correspondence, the relevant quantity is the entropy \emph{density} [see, e.g., \cite{kovtun}]. If the entropy density measures the information content of the planar black hole, then under Hawking evaporation\footnote{Large AdS black holes tend to achieve thermal equilibrium with their own Hawking radiation. Nevertheless they can evaporate if we change the boundary conditions. This can be achieved by coupling the boundary CFT with an auxiliary system \cite{aux}.} it would eventually reach the Page time \cite{page1, page2}, at which point firewall \cite{amps1, amps2} is argued to set in. [See also \cite{BPZ}]. However, if the true information content is given by the CR-volume, which is infinite, then presumably the Page time will never be reached for a planar black hole. See also, Sec.(3.3) of \cite{1412.8366} for related discussions.
[Of course, having an infinite capacity does not necessarily mean that it has that much information, but only that \emph{in principle, in could}.] 
Likewise, even for other black holes, depending on whether the area or the CR-volume encodes information, the Page time will be different.
The role of the CR-volume may therefore have profound implication on the onset of the Page time, as well as the firewall paradox. 

We would like to emphasize that as far as information loss is concerned, we must know what happens to the singularity in a theory of quantum gravity [see \cite{1412.8366} for further discussion on this emphasis]. This is because despite the large interiors, the worldlines of in-falling objects always hit the curvature singularities in finite proper time. In 4-dimensions, for asymptotically flat Schwarzschild black hole the maximal in-fall time is $\pi M$; for AdS neutral flat toral black hole, it is $\pi L/3$, as we have shown in Eq.(\ref{wow}). That is, classically the information crashes into the singularity and may still be lost. In addition, we did not discuss possible instabilities of the black hole solutions in this work. If we are really interested in information loss paradox, the final fate of a given black hole is an important issue --- if the black hole becomes unstable and evolves into another geometry, its large interior may be irrelevant to information loss. 

Lastly, we note that AdS black holes with flat horizons, especially the planar ones, are especially important in the various applications of AdS/CFT correspondence, e.g., to condensed matter and heavy ion collider physics [see \cite{1403.4886} and the references therein]. 
Since the field theory physics is unitary, this provides a strong reason to believe that the bulk physics is also unitary. Understanding how information loss paradox is solved in the bulk however remains an open problem.  There are recent attempts to probe the interiors of black holes in string theory \cite{1310.6334, 1412.1084}. If the CR-volume stores information, it would be important to understand its role in the AdS/CFT context.

%%%%%%%%%%%%%%%%%%%%%%%%%%%%%%%%%%%%%%%%%%%%%%%%%%%%%%%%%%%%%%%%%%%%%%%%%%%%%%%%%%%%%%%%%%%%%%%%%%%%%%%%%%%%%%%%%%%%%%%%%%%%%%%%%%%%%%%%%%%%%%%%%%%%%%%%%%

\addtocounter{section}{1}
\section*{\large{\textsf{Acknowledgement}}}
The author is grateful to Ingemar Bengtsson and Brett McInnes for comments and discussions.

%%%%%%%%%%%%%%%%%%%%%%%%%%%%%%%%%%%%%%%%%%%%%%%%%%%%%%%%%%%%%%%%%%%%%%%%%%%%%%%%%%%%%%%%%%%%%%%%%%%%%%%%%%%%%%%%%%%%%%%%%%%%%%%%%%%%%%%%%%%%%%%%%%%%%%%%%%

\end{document}